\documentclass[%
 reprint,
superscriptaddress,
 amsmath,amssymb,
 aps,
]{revtex4-1}
\usepackage{textcomp}
\usepackage{comment}
\usepackage{graphicx}
\usepackage{dcolumn}
\usepackage{bm}
\usepackage[utf8]{inputenc}
\usepackage{ragged2e}
\usepackage{subfig}
\usepackage{xcolor}
\usepackage{verbatim}
\usepackage{xr-hyper}
\usepackage{hyperref}
\usepackage[normalem]{ulem}

  \hypersetup{
  colorlinks = true,
  linkcolor = purple,
  citecolor = red,
urlcolor = blue!75!black
  }

\begin{document}

\hyphenation{pro-ba-bi-li-ties}
\hyphenation{mo-du-la-tion}
\hyphenation{stu-dies}
\hyphenation{pho-to-io-ni-za-tion}
\hyphenation{using}
\hyphenation{qual-i-ta-tive-ly}
\hyphenation{analy-sis}
\hyphenation{Engkvist}

\preprint{APS/123-QED}

\title{Attosecond photoionization dynamics in the vicinity of the Cooper minima in argon}

\author{C.~Alexandridi}
\affiliation{Universit\'e Paris-Saclay, CEA, LIDYL, 91191, Gif-sur-Yvette, France}
\author{D.~Platzer}
\affiliation{Universit\'e Paris-Saclay, CEA, LIDYL, 91191, Gif-sur-Yvette, France}
\author{L.~Barreau}
\affiliation{Universit\'e Paris-Saclay, CEA, LIDYL, 91191, Gif-sur-Yvette, France}
\author{D.~Busto}
\affiliation{Department of Physics, Lund University, Box 118, SE-221 00 Lund, Sweden}
\author{S.~Zhong}
\affiliation{Department of Physics, Lund University, Box 118, SE-221 00 Lund, Sweden}
\author{M.~Turconi}
\affiliation{Universit\'e Paris-Saclay, CEA, LIDYL, 91191, Gif-sur-Yvette, France}
\author{L.~Neori\v{c}i\'{c}}
\affiliation{Department of Physics, Lund University, Box 118, SE-221 00 Lund, Sweden}
\author{H.~Laurell}
\affiliation{Department of Physics, Lund University, Box 118, SE-221 00 Lund, Sweden}
\author{C. L.~Arnold}
\affiliation{Department of Physics, Lund University, Box 118, SE-221 00 Lund, Sweden}
\author{A.~Borot}
\affiliation{Universit\'e Paris-Saclay, CEA, LIDYL, 91191, Gif-sur-Yvette, France}
\author{J.-F.~Hergott}
\affiliation{Universit\'e Paris-Saclay, CEA, LIDYL, 91191, Gif-sur-Yvette, France}
\author{O.~Tcherbakoff}
\affiliation{Universit\'e Paris-Saclay, CEA, LIDYL, 91191, Gif-sur-Yvette, France}
\author{M.~Lejman}
\affiliation{Universit\'e Paris-Saclay, CEA, LIDYL, 91191, Gif-sur-Yvette, France}
\author{M.~Gisselbrecht}
\affiliation{Department of Physics, Lund University, Box 118, SE-221 00 Lund, Sweden}
\author{E.~Lindroth}
\affiliation{Department of Physics, Stockholm University, AlbaNova University Center, SE-106 91 Stockholm, Sweden}
\author{A.~L'Huillier}
\affiliation{Department of Physics, Lund University, Box 118, SE-221 00 Lund, Sweden}
\author{J.~M.~Dahlstr\"om}
\affiliation{Department of Physics, Lund University, Box 118, SE-221 00 Lund, Sweden}
\author{P.~Sali\`eres}
\affiliation{Universit\'e Paris-Saclay, CEA, LIDYL, 91191, Gif-sur-Yvette, France}

\date{\today}

\begin{abstract}
Using a spectrally resolved electron interferometry technique, we measure photoionization time delays between the $3s$ and $3p$ subshells of argon over a large 34-eV energy range covering the Cooper minima in both subshells. 
The observed strong variations of the $3s-3p$ delay difference, including a sign change, are well reproduced by theoretical calculations using the Two-Photon Two-Color Random Phase Approximation with Exchange. Strong shake-up channels lead to photoelectrons spectrally overlapping with those emitted from the $3s$ subshell. These channels need to be included in our analysis to reproduce the experimental data. Our measurements provide a stringent test for multielectronic theoretical models aiming at an accurate description of inter-channel correlation. 
\end{abstract}
\maketitle


Half a century after their theoretical description by L.~Eisenbud~\cite{Eisenbud}, E.~P.~Wigner~\cite{WignerPR1955} and F.~T.~Smith~\cite{SmithPR1960}, scattering delays ---also called Wigner delays--- can now be measured using attosecond spectroscopy, which allows for detailed studies of the correlated interactions within various quantum systems.
In practice, this is done by recording the spectral variation of the quantum phase of electron wave packets photo-emitted from solids~\cite{CavalieriNature2007,OssianderNature2018}, molecules~\cite{HaesslerPRA2009,HuppertPRL2016,BeaulieuScience2017,VosScience2018}, or atoms~\cite{SchultzeScience2010,KlunderPRL2011,GuenotPRA2012,GuenotJPB2014,PalatchiJPB2014,SabbarPRL2015,JordanPRA2017,OssianderNatPhys2017,IsingerScience2017,CirelliNC2018}.
Of particular interest in scattering physics are spectral structures in the continua of atoms/molecules such as autoionizing resonances~\cite{Beutler1935,FanoPR1961}, shape resonances \cite{DehmerJCP1972} and Cooper minima~\cite{CooperPR1962}, as they carry detailed information on the internal structure, electronic correlations, potential and orbital shapes. The rapidly varying phase in the vicinity of these structures has recently been investigated, e.g. for autoionizing~\cite{KoturNatComm2016,GrusonScience2016,BustoJPB2018,BarreauPRL2019,Turconi_2020} and shape resonances~\cite{HuppertPRL2016}.
As for Cooper Minima (CM), very few measurements have been performed, either in photoionization~\cite{PalatchiJPB2014} or photorecombination~\cite{SchounPRL2014,scarboroughAppSci2018} spectroscopy.

\begin{figure}
    \centering
    \includegraphics[width=0.99\linewidth]{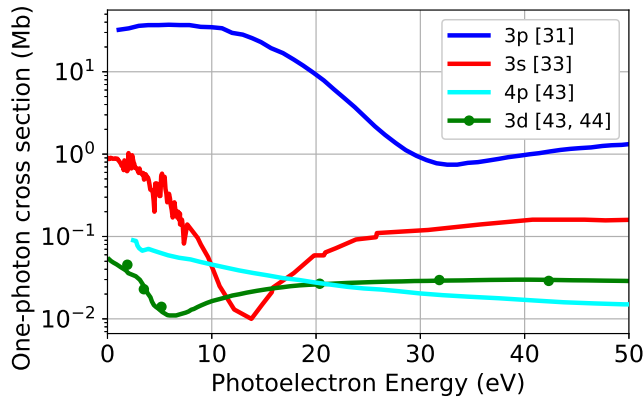}
    \caption{One-photon ionization cross sections of the $3p$ (blue) and $3s$ (red)  channels~\cite{MobusPRA1993,SamsonJElecSpec2002}, together with that of the satellite shake-up states $3s^23p^4(^1D)4p(^2P)$ (cyan)~\cite{1989WijesunderaPRA} and $3s^23p^4(^1D)3d(^2S)$ (green dots for experiments~\cite{1988BeckerPRL} and solid line for theory~\cite{1989WijesunderaPRA}) labelled $4p$ and $3d$ respectively. The ionization thresholds for $3p$, $3s$, $4p$ and $3d$ are $15.76$~eV, $29.24$~eV, $37.15$~eV and $38.60$~eV respectively.}
    \label{fig:example}
\end{figure}

Ionization of argon from the $n=3$ shell has attracted considerable attention due to abundant signatures of intra- and inter-orbital electronic correlations. In the case of the $3p$ subshell, intra-orbital correlation is important close to the ionization threshold, due to the so-called ground-state correlation~\cite{starace_theory_1982}. At higher photon energies, 
where the photoionization process can be described using single active electron models, 
the sign change in the $3p \rightarrow \varepsilon d$ radial transition matrix element leads to a CM close to 53~eV~\cite{CooperPR1962,SamsonJElecSpec2002}.
As for the $3s$ subshell, inter-orbital correlation is important, since the $3p \rightarrow \varepsilon d$ process is strongly coupled to $3s \rightarrow \varepsilon p$~\cite{AmusiaPLA1972}. This leads to a correlation-induced ``replica'' of the $3p$ CM in the $3s$ ionization channel, close to 42-eV photon energy, as shown in Fig.~\ref{fig:example}~\cite{MobusPRA1993}.

All these correlation effects are expected to leave an imprint on the scattering/photoionization delays between the $3s$ and $3p$ electrons, which motivated a large number of calculations during the past decade in the demanding region above the $3s$ threshold~\cite{KlunderPRL2011,GuenotPRA2012,KheifetsPRA2013,DahlstromPRA2012,DahlstromJPB2014,MagrakvelidzePRA2015,app2018Pi,BrayPRA2018,Sato2018,VinbladhPRA2019}. 
The methods qualitatively agree on the behavior of the $3p$ atomic delays, which are slightly negative over a large energy region around the $3p$ CM.
However, for the $3s$ case, the atomic delays close to the $3s$ CM strongly differ in magnitude and in sign depending on the degree of correlation included~\cite{DahlstromJPB2014,VinbladhPRA2019}. 
Up to now, two experiments using the RABBIT (Reconstruction of Attosecond Beating By Interference of two-photon Transitions) technique have aimed at measuring the photoionization time-delay difference in the $n=3$ shell of argon \cite{KlunderPRL2011,GuenotPRA2012}. Unfortunately, for experimental reasons, the results were limited to the 34--40~eV photon energy range, below the $3s$ and $3p$ CM, preventing a detailed comparison over the entire spectral region. 

In this Letter, we measure photoionization time delays between the $3s$ and $3p$ subshells of argon over a large energy range (34--68~eV) covering the CM in both subshells. The presence of multiple ionization channels leads to spectral congestion, which is one of the main experimental challenges. The spectrally resolved interferometric Rainbow RABBIT technique~\cite{GrusonScience2016,BustoJPB2018} allows us to substantially overcome this difficulty.
The experimental results show strong variations of the $3s-3p$ delay difference, with a change of sign in the $3s$ CM region. They are compared to theoretical calculations using the Two-Photon Two-Color Random Phase Approximation with Exchange (2P2C-RPAE) method~\cite{VinbladhPRA2019}. The agreement is excellent in ---and above--- the $3p$ CM region (45--68~eV), and unsatisfactory in the $3s$ CM region (34--45~eV). We show that the presence of electrons from strong shake-up (SU) channels overlapping with $3s$ electrons is the likely reason for this deviation. By taking into account the dominant SU contribution, a good agreement between theory and experiment can be obtained in the 34--39~eV range. 

The experiments were performed independently at the ATTOLab facility in Saclay, France and at Lund University, Sweden. The details of the two setups are described in the Supplemental Material \cite{*[{See Supplemental Material at [url] for details on the experimental and theoretical methods}] [{}] SM}. Briefly, intense infrared (IR) femtosecond pulses are split in a Mach-Zehnder interferometer. In one arm of the interferometer, high-order harmonics are generated in neon gas and spectrally filtered by metallic foils. In the other arm, a small fraction of the IR radiation is temporally delayed. Both beams are recombined and then focused into an argon gas jet. The emitted electrons are detected with a 2\,m-long magnetic bottle electron spectrometer (MBES).   

\begin{figure}
\centering
\includegraphics[width=0.99\linewidth]{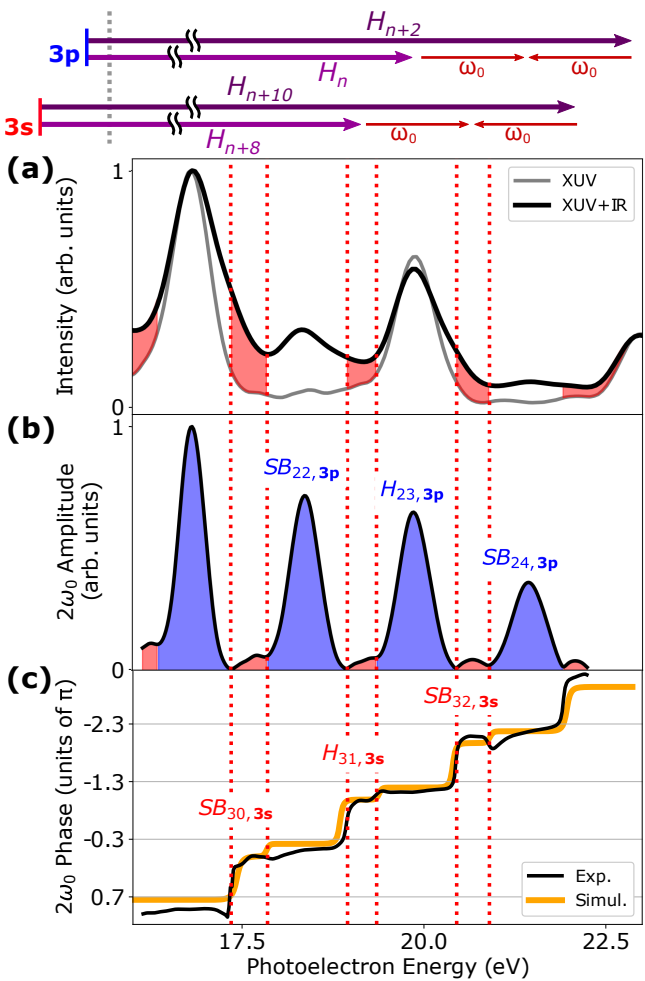}
\caption{(a) XUV+IR photoelectron spectrum (black), obtained in Saclay by integrating the RABBIT spectrogram over the delay, as compared to the XUV-only spectrum (gray), and corresponding multiphoton transitions for the $3s$ and $3p$ channels. Spectral amplitude (b) and phase (c) of the $2\omega_0$ oscillations obtained from a Fourier transform at each energy of the RABBIT spectrogram. In (b), the amplitudes related to the $3p$ ($3s$) channel are highlighted in blue (red), respectively. The red dashed lines indicate the spectral regions with dominant $3s$ contributions. In (c), the measured phase evolution (black line) is compared to the simulated one (orange line) \cite{SM}. }
\label{fig:AmpPhase}
\end{figure}

To measure the photoionization time delays, we used the Rainbow RABBIT technique~\cite{GrusonScience2016} which is schematically illustrated in Fig.~\ref{fig:AmpPhase} (top). A comb of coherent harmonics ionizes argon atoms, creating one-photon electron wave packets (EWP) from both $3s$ and $3p$ subshells. The corresponding electron peaks are referred to as ``$3s$ or $3p$ harmonics'' in the following. By adding the weak IR dressing field ($\sim$ $10^{11}$~W/cm$^2$), replicas of the initial EWPs are created by two-photon XUV$\pm$IR transitions to the same final states. Their interference gives rise to the so-called sidebands (SB), the intensity of which oscillates as a function of the delay $\tau$ between the XUV and IR pulses as:
\begin{equation}
\begin{split}
S\!B_{n,i}(\tau)&=A_{n,i}+ B_{n,i}\cos[2\omega_0\tau - \Delta\phi_{n}^{\mathrm{XUV}} -\Delta\phi_{n,i}^\mathrm{A}]
\end{split}
\label{Eq:SB}
\end{equation}
where $i$ is the ionization channel ($3s$, $3p$,~etc.), $\omega_0$ is the angular frequency of the driving laser, $\Delta\phi_{n}^{\mathrm{XUV}} = \phi_{n+1}-\phi_{n-1}$ is the phase difference between two consecutive harmonics with orders $n\pm1$~\cite{MairesseScience2003} and $\Delta\phi_{n,i}^\mathrm{A}$ is the phase difference between the two-photon transition dipole matrix elements. In the so-called asymptotic approximation \cite{DahlstromChemPhys2013}, $\Delta\phi_{n,i}^\mathrm{A}$  can be expressed as the sum of two contributions, $\Delta\eta_{n,i}+\Delta\phi_{n,i}^\mathrm{cc}$, where $\eta_{n\pm 1,i}$ is the scattering phase accumulated by the EWP in the one-photon (XUV) transition, which is intrinsic to the target atom, and $\phi_{n\pm 1,i}^\mathrm{cc}$ is a quasi-universal measurement-induced phase shift due to the electron being probed by the IR laser field in a long-range potential with a Coulomb tail \cite{DahlstromChemPhys2013}.

Group delays can be defined through $\tau \approx \Delta\phi/2\omega_0$.
The measured delay can be expressed as $\tau_{n}^\mathrm{XUV}+\tau_{n,i}^\mathrm{A}$. Introducing the Wigner delay  $\tau_{n,i}^\mathrm{W}\approx\Delta \eta_{n,i}/2\omega_0$,  $\tau_{n,i}^\mathrm{A}\approx \tau_{n,i}^\mathrm{W} + \tau_{n,i}^\mathrm{cc}$. 
Since the $3s$ and $3p$ photoelectrons are ionized by the same harmonic comb, the $\tau_n^\mathrm{XUV}$ contribution of the ionizing radiation can be removed by calculating the difference of the delays for the two channels, giving direct access to $\tau_{3s}^{\mathrm{A}} - \tau_{3p}^{\mathrm{A}}$ with high accuracy.

Two main experimental difficulties have prevented the measurement of this delay difference over the whole CM region. The first is that the ionization cross section is much weaker for $3s$ than for $3p$, as shown in Fig.~\ref{fig:example}. The second is the spectral overlap of the two channels: the $3s$ sidebands $S\!B_{n,3s}$  fall only 0.47~eV from the much more intense $3p$ harmonics $H_{n-9,3p}$ due to the difference in ionization energies $\Delta E_{3s-3p} = 13.48$~eV $= 9\hbar\omega_0 - 0.47$~eV. 
To cope with these difficulties, previous studies~\cite{KlunderPRL2011,GuenotPRA2012} spectrally isolated the two contributions by photoionizing with only four harmonics ($H_{21}$--$H_{27}$), selected by using a combination of filters and generation in argon.
The main drawback of this approach is that it constrains the usable energy range to a 10-eV window below 40~eV.

In the present work, harmonics generated in neon were used as the ionizing radiation. Neon has a  smaller generation efficiency than argon, but exhibits a quite flat harmonic spectrum with a much higher cut-off energy. In combination with a single 200~nm-thick Al filter, a broad spectrum (20--72~eV) including harmonics $H_{13}$--$H_{45}$ is obtained. In these conditions, the $3p$ harmonics overshadow the $3s$ sidebands. In Fig.~\ref{fig:AmpPhase}(a), the contributions of the $3s$ SBs (harmonics) in the XUV+IR spectrum appear as small shoulders highlighted on the blue side of the $3p$ harmonics (SBs), as evidenced by the comparison with the XUV-only spectrum. 
In order to separate the two contributions, we perform a Rainbow RABBIT analysis of the recorded spectrogram, \textit{i.e.} we analyze the $2\omega_0$ oscillations for each energy $E$ in the spectrum (see Eq.~(\ref{Eq:SB})). 
The amplitude and phase of these oscillations are shown in Fig.~\ref{fig:AmpPhase}(b) and (c) respectively.

\begin{figure}
\centering
\includegraphics[width=9cm]{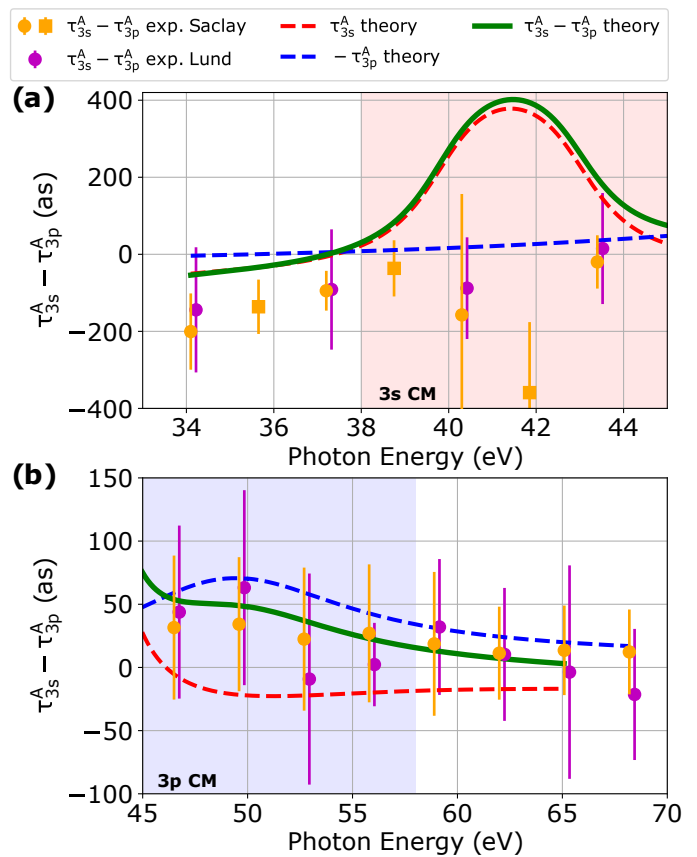}
\caption{Atomic delay difference between $3s$ and $3p$ ionization channels in the vicinity of the $3s$ CM (a) and $3p$ CM (b), highlighted in red and blue respectively: experimental data from Lund (magenta dots) and Saclay (orange markers, circles for sidebands, squares for harmonics) together with simulations using the 2P2C-RPAE model (green line). The simulations for the $3s$ (red-dashed line) and $3p$ (blue-dashed line) delays are also shown.} 
\label{fig:exp_vs_theory}
\end{figure}

Between consecutive harmonics and sidebands of the same ionization channel, a dephasing of $\simeq\pi$ is expected, as a result of the conservation of the total number of electrons~\cite{SM}.
Fig.~\ref{fig:AmpPhase}(c) shows multiple phase jumps, some of them close to $\pi$, others much less than $\pi$, which occur between sidebands (or harmonics) of different subshells. 
In addition to the high resolution of the MBES, the combination of amplitude and phase measurements allows us to distinguish the $3s$ contribution from the strong neighboring $3p$ electron peaks (see the dotted lines).

The measured $\tau_{3s}^{\mathrm{A}} - \tau_{3p}^{\mathrm{A}}$ delay differences are plotted in Fig.~\ref{fig:exp_vs_theory}. The results reveal an interesting feature, namely a change of sign around 43~eV.
The negative sign in the low energy region (close to the $3s$ threshold) is in part explained by the large negative delay contribution due to continuum--continuum transitions of slow photoelectrons~\cite{DahlstromJPB2012}. 
However, in neon, the atomic delay difference between the $2s$ and $2p$ electrons stays negative up to high $\sim$100-eV~energies~\cite{IsingerScience2017}. The positive sign in argon is thus a probable signature of the CM. 
Finally, the delay difference converges towards zero for higher energies where the $3s$ and $3p$ cross sections are unstructured.

The low energy region is a particularly difficult range for the measurements because of the very weak signal of the $3s$ channel due to the CM, resulting in more than three orders of magnitude difference with the $3p$ channel (see Fig.~\ref{fig:example}(a)). An extended analysis over a larger set of data was thus performed to improve the statistical significance of the result in the 34--43~eV spectral range corresponding to orders 22 to 27. There, the phases were extracted for both sideband and harmonic peaks. 
As discussed in Fig.~\ref{fig:AmpPhase}(b), the harmonic phase are shifted by $\sim\pi$ relative to the SBs. However, this $\pi$ phase-shift cancels when we calculate the delay \textit{difference} between $3s$ and $3p$ harmonics. As shown in Fig.~\ref{fig:exp_vs_theory}, the harmonic delays behave as an average of the neighboring sideband delays \cite{SM}, with the only exception being $H_{27}$ at 42~eV.

In Fig.~\ref{fig:exp_vs_theory}, the experimental results are compared with theoretical predictions using the recently developed 2P2C-RPAE method~\cite{VinbladhPRA2019}. 
It consists in calculating complete self-consistent two-photon processes, including electron correlation in bound--continuum and continuum--continuum transitions, ion polarization effects and the reversed photon time orders. 
The orbital energies of $3p$ and $3s$ are adjusted to fit the experimental ionization thresholds and the interaction with the fields is computed in the length gauge. 
The calculated $\tau_{3p}^{\mathrm{A}}$ and $\tau_{3s}^{\mathrm{A}}$ include the contribution from all emission angles, which corresponds to our experimental configurations. For $\tau_{3p}^{\mathrm{A}}$, this integration significantly modifies the delays as compared to that in the XUV polarization direction~\cite{DahlstromJPB2014,PalatchiJPB2014}, while for $\tau_{3s}^{\mathrm{A}}$ little effect is observed due to the single $3s \rightarrow \varepsilon p$ transition. 
Earlier work~\cite{DahlstromJPB2014} shows that the full one-photon RPAE theory yields a $3s$ CM at 40~eV, while the restricted one-photon RPAE calculation with only intrashell ($n=3$) correlation produces a CM close to 42~eV in better agreement with one-photon cross section measurements \cite{MobusPRA1993}. For this reason, we use the latter method  for the $3s$ 2P2C-RPAE simulations in Fig.~\ref{fig:exp_vs_theory}, which moves the positive peak of $\tau_{3s}^{\mathrm{A}}$ from 40~eV to the correct position of 42~eV \cite{SM}. 
 
An excellent agreement between the simulations and the experimental data is observed in the $3p$ CM region and above (45--68~eV range), as shown in Fig.~\ref{fig:exp_vs_theory}~(b). 
For the lower energies shown in Fig.~\ref{fig:exp_vs_theory}~(a), a small deviation in the 34--39~eV range and a disagreement in the 39--42~eV range can be noticed. 

\begin{figure}
    \centering
    \includegraphics[width=\linewidth]{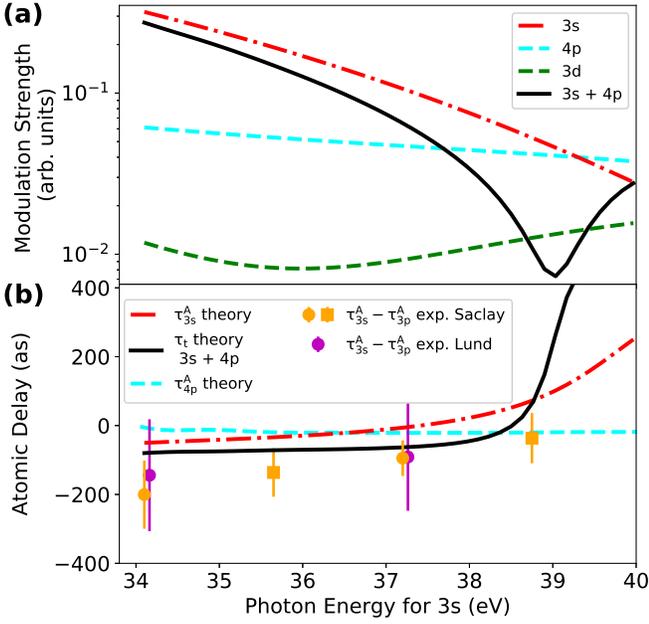}    
    \caption{(a)~Estimated modulation strength $B_{n,i}$ of the RABBIT signal for the $3s$ (red), $4p$ (cyan) and $3d$ (green) channels, and effective modulation strength $B_t$ of the incoherently combined $3s$ and $4p$ contributions, denoted $3s+4p$ (black). (b)~Theoretical atomic delays for the $3s$ (red) and $4p$ (cyan) channels, and effective atomic delay for the incoherent $3s+4p$ case (black). The experimental points are plotted again for comparison (in this region, $\tau^{\mathrm{A}}_{3p}\sim 0$).
    }
    \label{fig:fig4}
\end{figure}
We attribute these differences to the presence of unresolved additional ionic channels that perturb the measurements. In this $3s$ CM region where the cross section drops, two shake-up (SU) channels associated with the $3s^23p^4(^1D)3d(^2S)$ and $3s^23p^4(^1D)4p(^2P)$ ionic states, hereafter denoted $3d$ and $4p$ channels, give significant contributions to the photoelectron spectrum and may even dominate over $3s$, as shown in Fig.~\ref{fig:example}.
The $4p$ channel has $H_{n+5,4p}$ overlapping with $S\!B_{n,3s}$ due to the difference in ion excitation energies being five times the laser photon energy, $\Delta E_{4p-3s}=7.90\,\mathrm{eV}\approx 5\hbar\omega_0$. Similarly, the $3d$ channel has $S\!B_{n+6,3d}$ overlapping with $S\!B_{n,3s}$, due to $\Delta E_{3d-3s}=9.34\,\mathrm{eV}\approx 6\hbar\omega_0$.

We first consider the 34--39~eV range, where the $3s$ and $4p$
are the dominant channels.
We model the RABBIT modulation amplitude for a given channel by assuming that it is proportional to the one-photon cross section at a given final kinetic energy of the photoelectron \cite{SM}.
This allows us to make simple estimates of the modulation strengths $B_{n,i}$ in Eq.~(\ref{Eq:SB}) assuming constant harmonic photon flux [see Fig.~\ref{fig:fig4}~(a)].
The crossing between the $3s$ and $4p$ modulation strengths close to 39~eV results in a significant SU contamination of the $3s$ time delay measurements in this energy range.

The incoherent sum of the two signals, $S\!B_{n,3s}$ and $H_{n+5,4p}$, results in a phase-shifted and amplitude-modified oscillating function  $B_t\cos(2\omega_0\tau - \theta_t) $, where: 
\begin{equation}
\theta_t=\tan^{-1}\left(\frac{\sin(\delta_n)}{B_{n,3s}/B_{n+5,4p}+\cos(\delta_n)}\right)+\theta_{n,3s}, \\
\label{eq:tantheta}
\end{equation}
\begin{equation*}
    B_t=\sqrt{(B_{n+5,4p})^2+(B_{n,3s})^2+2B_{n+5,4p}B_{n,3s}\cos(\delta_n)},
\end{equation*}
with $\theta_{n,3s}=\Delta\phi_n^\mathrm{XUV}+\Delta\phi_{n,3s}^\mathrm{A}$. The relative phase between the $4p$ and $3s$ oscillations, $\delta_n$, is defined by:
\begin{align}
\delta_n=\Delta\phi_{n+5}^{\mathrm{XUV}} - \Delta\phi_{n}^{\mathrm{XUV}} + \Delta\phi_{n+5,4p}^{\mathrm{A}} - \Delta\phi_{n,3s}^{\mathrm{A}} +\pi,
\label{eq:delta}
\end{align}
where the $\pi$ factor accounts for the opposite oscillations of $H_{n+5,4p}$ with respect to the neighboring $4p$ sidebands \cite{SM}. 
$\Delta\phi_{n+5}^{\mathrm{XUV}} - \Delta\phi_{n}^{\mathrm{XUV}} \approx0.14~\pi$
is determined by an independent measurement of the attosecond chirp.  
The $3s$ atomic phase is computed using the 2P2C-RPAE theory restricted to $n=3$ correlation, while the $4p$ atomic phase is computed using the asymptotic approximation, as described in \cite{SM}.
The resulting atomic delay for $4p$ SU, shown in Fig.~\ref{fig:fig4}~(b), is found to be small, due to cancellation of the positive $\Delta\phi_{n+5,4p}^\mathrm{W}$ and negative $\Delta \phi_{n+5,4p}^\mathrm{cc}$ contributions. This implies that the atomic phase, $\Delta\phi_{n+5,4p}^{\mathrm{A}}$, plays a negligible role in Eq.~\ref{eq:delta}. Similarly $\Delta\phi_{n,3s}^{\mathrm{A}}$, as shown in Fig.~\ref{fig:exp_vs_theory}(a), is small so that $\delta_n$ is close to $\pi$.

In the 34--38.5~eV range where $B_{n+5,4p}\ll B_{n,3s}$, $\theta_t \approx (\pi-\delta_n) B_{n+5,4p}/B_{n,3s}+\theta_{n,3s}$; The contribution of the $4p$ SU channel shifts down the effective atomic delay, in excellent agreement with the experimental results in Fig.~\ref{fig:fig4}~(b). Around 39~eV, the similar amplitudes and $\sim\!\!\pi$-shifted oscillations of the two channels result in a minimum of the total modulation strength $B_t$ and a fast variation of the effective delay as illustrated in Fig.~\ref{fig:fig4} by the black lines.
However, the $3d$ SU channel cannot be neglected anymore in this energy region.
This SU process is much more complicated to estimate because it originates from the highly correlated $3s$ channel and exhibits a Cooper-like minimum as shown in Fig.~\ref{fig:example}~(a) and \ref{fig:fig4}~(a). The 39--42~eV range is thus a transition region where at least three dephased incoherent channels contribute. This might explain why the delay measured at harmonic 27 (close to 42 eV), cannot be easily reproduced by modelling.

In conclusion, we have demonstrated the potential of the Rainbow RABBIT method to separate the contributions of the $3s$ and $3p$ ionization channels in argon and to measure the corresponding atomic delays for a wide range of energies (34--68~eV) that includes both $3s$ and $3p$ Cooper minima. The extracted $3s-3p$ delay differences are in good agreement with earlier results in the 34--40 eV range \cite{GuenotPRA2012}, being, however, at variance with more recent measurements \cite{Hammerland2019arxiv}. Our results are in excellent agreement with the predictions of many-body perturbation theory in a 24~eV range around the $3p$ Cooper minimum, revealing the high accuracy of both experiment and theory in this region. Furthermore, we identify two strong shake-up ionization channels, the contributions of which are probably responsible for the discrepancy observed in the $3s$ Cooper minimum region. This calls for further investigations to clarify their role. For instance, using a mid-IR driving wavelength from an optical parametric amplifier would allow for better sampling and tunability in order to separate the $3s$, $3p$ and shake-up channels. 
This study thus provides a new step towards improving our understanding of the complex nature of correlated multielectron ionization dynamics.

C.~A. and D.~P. contributed equally to this work. This research was supported by the ANR-15-CE30-0001- CIMBAAD, ANR-11-EQPX0005-ATTOLAB, ANR-10-LABX-0039-PALM, COST/CA18222-AttoChem Action and Laserlab-Europe EU-H2020-654148. The authors affiliated in Sweden acknowledge support from the Swedish Research Council (Grants 2018-03845, 2013-08185 and 2016-03789) and the Knut and Alice Wallenberg Foundation (Grant No. 2017.0104). J. M. D. acknowledges support from the Swedish Foundations' Starting Grant by the Olle Engkvist's Foundation. The Lund group thanks Raimund Feifel and Richard J.~Squibb, Gothenburg University, for borrowing their Magnetic Bottle Electron Spectrometer. 


%

\end{document}


\hyphenation{va-lues}

\title{Supplemental Material}
\maketitle
\section{Experimental methods}
\subsection{Experimental setup at the ATTOLab facility}
The Titanium:Sapphire laser of the ATTOLab facility at CEA-Saclay delivers 23-fs 13-mJ 800-nm pulses at 1~kHz repetition rate. The setup consists of a Mach-Zehnder interferometer where 90\% of the initial 7-mJ beam is focused with a $f=2$~m lens into a neon gas cell to generate an attosecond pulse train by high harmonic generation. An anti-reflective coated silica plate, set at grazing incidence (78.5\textdegree), is used to transmit most of the strong generating infrared (IR) beam and to reflect the generated extreme ultra-violet (XUV) light. After an aluminium filter used to block the remaining IR radiation, the attosecond pulse train is focused by a gold-coated toroidal mirror into an argon gas jet in the interaction region of a 2m-long magnetic bottle electron spectrometer (MBES).  The remaining 10\% of the initial beam (the dressing beam) enters a delay line with piezo-electric translation stage in order to control the XUV-IR delay with interferometric accuracy. It is then focused with a combination of two lenses of $f_1=1140$~mm and $f_2=400$~mm, and reflected on a drilled mirror ($d = 2$~mm) that transmits the XUV beam. The two beams are then spatially and temporally overlapped in the interaction region of the MBES.
 
Obviously, the high spectral resolution of the MBES is a strong asset for resolving the two ionization channels. Since its resolution decreases with electron energy, a retarding potential was used to shift the sidebands of interest towards lower kinetic energies. This way, the spectral phases inside the $3s$ and $3p$ sideband peaks are quite flat without any fast variations apart from a linear slope that could be attributed to the harmonic blue shift \cite{BustoJPB2018}. This does not affect the measurement and, to further improve the signal-to-noise ratio for the $3s$ channel, we integrate in energy over typically 400~meV before extracting the $2\omega_0$ phase.
 
Using different retarding potentials, we were able to measure the phases of the $3s$ and $3p$ sidebands up to 68~eV. For photon energies around 60--68~eV, the signal-to-noise ratio becomes low due to the decrease in ionizing harmonic intensity. In order to ensure the accuracy of our results, additional measurements were performed by adding a 200-nm zirconium filter. The combined Zr-Al filter has a good transmission in a narrow energy window of 60--73~eV and allows the clear spectral separation of the $3s$ and $3p$ peaks. The phases of sidebands 40, 42 and 44 measured with high accuracy in these conditions were found very similar to the ones measured with only the Al filter. 

\subsection{Experimental setup at Lund University}
At Lund University, the Titanium:Sapphire laser delivers 40-fs 3.5-mJ 800-nm pulses at 1~kHz repetition rate. The setup consists of a Mach-Zehnder interferometer where 70\% of the IR beam is focused with a $f=50$~cm spherical mirror into a neon gas cell to generate attosecond pulse trains. An aluminium filter is used to block the remaining infrared radiation and to spectrally filter the XUV radiation. The attosecond pulse trains are focused by a gold-coated toroidal mirror into an argon gas jet in the interaction region of a 2m-long MBES. The remaining 30\% of the initial beam enters a delay line with a piezo-electric translation stage providing interferometric accuracy. The probe arm also includes a chopper wheel, which blocks the IR every other shot, as well as a spherical mirror with $f = 50$~cm. The two beams are recombined on a drilled mirror that transmits the XUV beam and reflects the dressing IR beam. The two beams are then spatially and temporally overlapped in the target gas jet.
 
The spectral resolution $\delta \!E$ of our MBES varies with the electron kinetic energy $E$ as $\sim E/\delta \!E=80$. In order to optimize the spectral resolution we apply a small retarding potential to slow down the photoelectrons. The high spectral resolution of our MBES allows us to distinguish between the $3s$ and $3p$ photoelectrons peaks over the whole energy range studied so that we do not vary the retarding potential. The weak $3s$ sidebands are identified by subtraction of the XUV+IR and IR-only photoelectron spectra for each delay. To increase the signal-to noise ratio, the sideband signal is integrated over an energy region in which sideband phase is approximately constant and the averaged phase is extracted by fitting the oscillation of the integrated signal using a nonlinear Levenberg-Marquardt algorithm \cite{LevenbergQAM1944,MarquardtJIAM1963}.
 
\subsection{Statistical data analysis}
Each delay shown in Figs.~\ref{fig:exp_vs_theory} and \ref{fig:fig4} is the resulting value $\overline{x}_w$ of a weighted average over independent measurements $x_i$:
\begin{equation}
    \overline{x}_w = \dfrac{\sum\limits_{i=1}^N\dfrac{x_i}{\sigma_i^2}}{\sum\limits_{i=1}^N\dfrac{1}{\sigma_i^2}},
\end{equation}
with $\sigma_i$ being the standard deviation estimated for each single measurement. In the Lund data, $\sigma_i$ corresponds to the uncertainty of the cosine fit of the sideband oscillations used to extract the $2\omega_0$ phase.
In the Saclay data, the $2\omega_0$ phase is extracted by Fourier transform and $\sigma_i$ is estimated from the $2\omega_0$ peak signal to noise ratio (in the spectral domain).
To calculate the total error $\delta x$, we use the following formula:
\begin{equation}
    \delta x = \sqrt{\frac{N\sum\limits_{i=1}^N\dfrac{(x_i-\overline{x}_w)^2}{\sigma_i^2}}{(N-1)\sum\limits_{i=1}^N\dfrac{1}{\sigma_i^2}}+\dfrac{1}{\sum\limits_{i=1}^N\dfrac{1}{\sigma_i^2}}}
\end{equation}
The first term in the square root represents the weighted estimator of the variance. As the number of performed measurements was probably not enough for this term alone to be statistically relevant ($\sim$10 in Saclay, $\sim$4 in Lund), we added a second term which corresponds to the variance of the weighted mean that prevents $\delta x$ from being artificially low when the small number of data points happens to yield a small dispersion. This formula is in the spirit of Eq.~8 of \cite{Cattaneo_2016} except for the second term that here yields a decrease of $\delta x$ in $1/\sqrt{N}$.

\maketitle
 \section{Basic simulations of the RABBIT spectrogram} 
 In order to simulate the 2$\omega_0$ phase shown in Fig.~\ref{fig:AmpPhase}(b), a simple model was used. We are interested in simulating the RABBIT trace of the combination: $H_{21,3p}$ + $S\!B_{30,3s}$ + $S\!B_{22,3p}$+$H_{31,3s}$+$H_{23,3p}$+$S\!B_{32,3s}$+$S\!B_{24,3p}$+$H_{33,3s}$.\\

We first describe the sidebands as:
\begin{align}
\begin{split}
S\!B_{n,i}(\tau) =& \left |M_{n-1,i}\right |^2 + \left |M_{n+1,i}\right |^2 \\ &+ 
2\left | M_{n-1,i} \right | \left | M_{n+1,i} \right | \\ &\times \cos(2\omega_0 \tau - \Delta \phi_{n}^\mathrm{XUV} - \Delta \phi_{n,i}^\mathrm{A})
\end{split}
\label{eq:SB_1}
\end{align}

where $i$ stands for the $p$ or $s$ channel, $n$ for the different orders, $\Delta \phi_{n}^\mathrm{XUV}$ for the phase difference between the neighboring harmonics, and $\Delta \phi_{n,i}^\mathrm{A}$ for the phase difference between the two-photon transition dipole matrix elements $M_{n\pm1,i}$. 

For estimating $\Delta \phi_{n}^\mathrm{XUV}$, we use the attochirp value of 0.0577~rad/eV extracted from a measurement performed in ATTOLab in the same generation conditions but with neon as target gas, so that the corresponding atomic phase variation is negligible. This value  agrees well with the attochirp measured at Lund University, $0.0442\pm0.0164$~rad/eV.
For $\Delta \phi_{n,i}^\mathrm{A}$, we use the values calculated using the 2P2C-RPAE model, corresponding to the delays shown in Fig.~\ref{fig:exp_vs_theory}. 

We approximate the amplitude of the two-photon transition dipole matrix elements by $|M_{n\pm1,i}| \approx |\mathcal{E}_{n\pm1}| \sqrt{\sigma_{n\pm1,i}}$, i.e., the product of the amplitude of the  harmonic fields by the square root of the corresponding one-photon ionization cross-sections given by \citep{MobusPRA1993} for the $3s$ and  \citep{SamsonPRL1974} for the $3p$ channel.

Simulating the oscillating signal of a harmonic peak requires some additional approximations as detailed in~\citep{Ruchon2013}.
In a first-order approximation, a harmonic peak will be modulated due to the loss of electrons that are transferred to the neighboring sidebands as:
\begin{equation}
\begin{split}
H_{n+1,i}(\tau) \approx & \ \alpha \left |M_{n+1,i}\right |^2 - S\!B_{n,i}(\tau)\\
&- S\!B_{n+2,i}(\tau),
\end{split}
\label{eq:HH}
\end{equation}
where $\alpha$ accounts for the difference in intensity of the harmonic and SB peaks
(1-photon with respect to 2-photon transitions, typically a factor 10).
Indeed, the dressing field in RABBIT is a weak perturbation inducing 2-photon XUV+IR transitions that redistribute electrons between the harmonic and sideband peaks depending on the XUV-IR delay. Note that, in the case of slowly-varying cross-sections and harmonic fields of similar amplitudes, the harmonic peak intensity can be simplified to:
\begin{widetext}
\begin{equation}
\begin{split}
H_{n+1,i}(\tau) =& \ (\alpha-4) \left | M_{n+1,i} \right |^2
-4  \left | M_{n+1,i}\right |^2 
 \times \cos \left( 2\omega_0\tau-\frac{1}{2}\left[\Delta \phi_{n}^\mathrm{XUV} +\Delta \phi_{n+2}^\mathrm{XUV} + \Delta \phi_{n,i}^\mathrm{A} + \Delta \phi_{n+2,i}^\mathrm{A}\right] \right) \\
&\times \cos\left(\frac{1}{2}\left[\Delta \phi_{n}^\mathrm{XUV} -\Delta \phi_{n+2}^\mathrm{XUV} + \Delta \phi_{n,i}^\mathrm{A} - \Delta \phi_{n+2,i}^\mathrm{A}\right] \right).
\end{split}
\label{eq:harm_ready}
\end{equation}
\end{widetext}

Harmonic $H_{n+1,i}$ will then oscillate with a total phase equal to $\pi+(\theta_{n,i}+\theta_{n+2,i})/2$, where $\theta_{n,i}$ ($\theta_{n+2,i}$) is the total phase of $S\!B_{n,i}$ ($S\!B_{n+2,i}$), resp. In regions of slow variation of $\theta_{n,i}$, the harmonic orders are thus dephased by $\sim\pi$ with respect to the neighboring sidebands. 

\begin{figure}
\centering
\includegraphics[width=7cm]{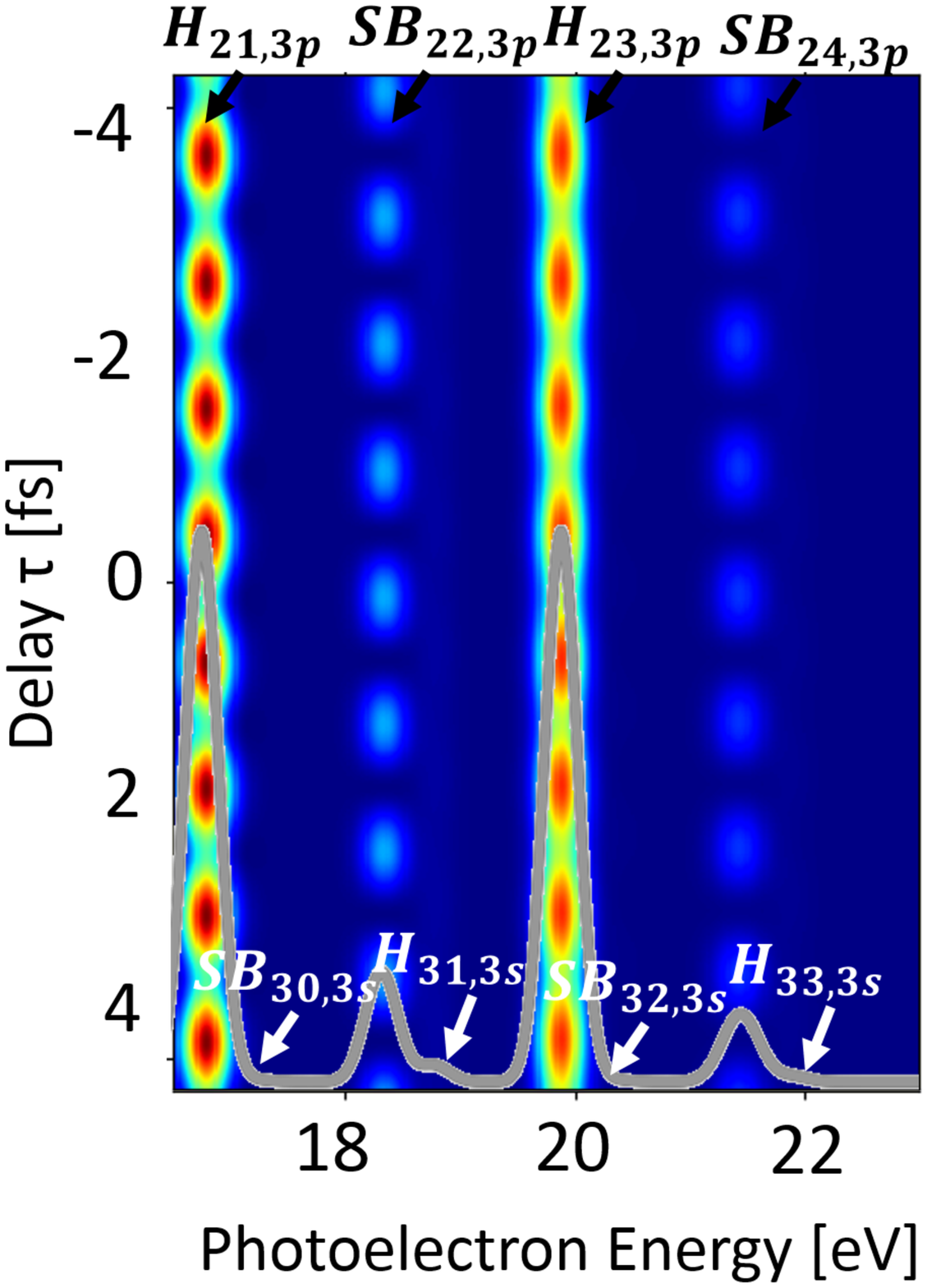}
\caption{RABBIT spectrogram simulated using the model described in the text and corresponding to the phase data shown in Fig.~\ref{fig:AmpPhase}(b).} 
\label{fig:SM_rabbit}
\end{figure}

After convolving the oscillating signals of the  sidebands and harmonics with the spectrometer response function $f_{sp}$ for each delay, we add them incoherently in order to obtain the total RABBIT spectrogram of Fig.~\ref{fig:SM_rabbit}.
For this reconstruction, we used Gaussian harmonic intensities with 300-meV FWHM and a Gaussian spectrometer response with 100-meV FWHM, values that fitted best our experimental results, together with $\alpha=14$. After applying the same Rainbow RABBIT analysis as for the experimental measurements, the $2\omega_0$ phase, shown in Fig.~\ref{fig:AmpPhase}(b), and the corresponding $2\omega_0$ amplitude were extracted. 
In this spectral region of slow variation of $\Delta \phi_i^\mathrm{A}$, the neighboring harmonic and sideband peaks of a given channel oscillate in opposite phase, as clearly seen in Fig.~\ref{fig:SM_rabbit} and revealed by the Rainbow RABBIT analysis.

\maketitle
\section{Correlation effects at the $3s$ Cooper minimum}
\begin{figure}[h!]
\centering
\includegraphics[width=\linewidth]{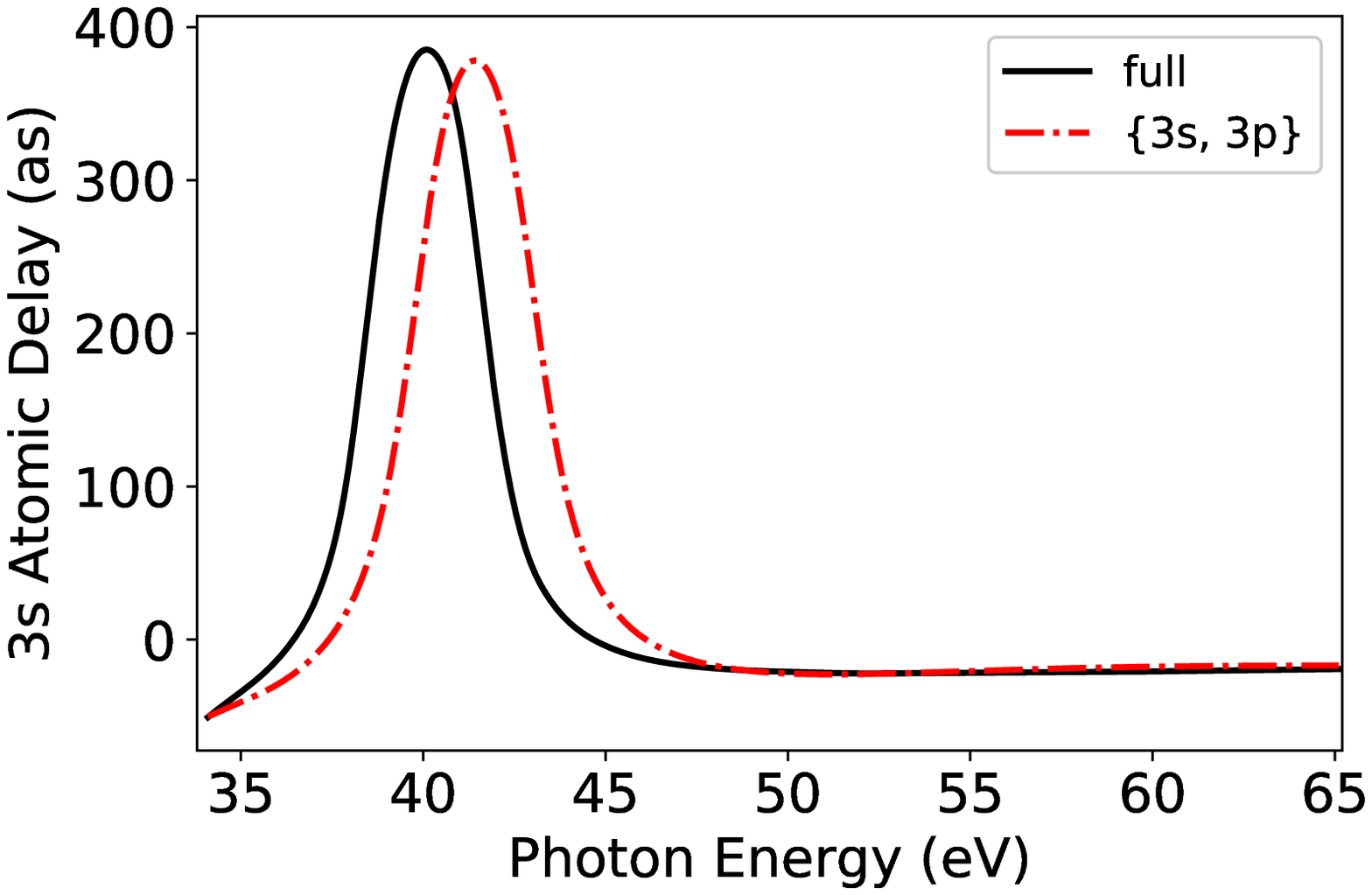}
\caption{Comparison of the atomic delays for $3s$ using the {\it full}~2P2C-RPAE and $\{3s,3p\}$~2P2C-RPAE theories.} 
\label{fig:comp_3s}
\end{figure}

Fig.~\ref{fig:comp_3s} contains a comparison of the atomic delays for $3s$ computed using the {\it full}~2P2C-RPAE theory, which includes inter-orbital correlation between all argon orbitals $\{1s,2s,2p,3s,3p\}$ with that of a restricted 2P2C-RPAE calculation, which includes only $M$-shell correlation $\{3s,3p\}$. The atomic delay of $3s$ is shifted up in the restricted calculation, which leads to better agreement with the experimental results. This shows that the atomic delay is sensitive to the electron correlation model close to the $3s$ CM, but that the corrected position of the $3s$ CM with the restricted 2P2C-RPAE model is insufficient to explain the deviation from the experimental measurements at low energy (34--39~eV).

\section{Shake-up atomic delay}
The $4p$ shake-up is modelled as an intermediate photoionization process, described by one-photon RPAE from $3p$, with a modification to enforce energy conservation as the ion is excited (implemented as an increase of the binding energy, $\Delta E_{4p-3p}$, in the final propagator). 
 The Wigner delay of $4p$ is computed in the forward direction, which is reasonable approximation for the photoionization from $3p$ at required energy range~\cite{PalatchiJPB2014}.
Accurate values of $\tau^\mathrm{cc}$ are taken from numerical calculations of noble gas atoms~\cite{DahlstromPRA2012}.

%